\documentclass[showpacs,preprintnumbers,amsmath,amssymb]{revtex4}
\usepackage{amssymb}
\usepackage{amsmath}

\usepackage{graphicx}
\usepackage{dcolumn}
\usepackage{bm}
\usepackage{graphics}

\begin{document}


\title{Scattering of a relativistic scalar particle by a
cusp potential}
\author{V\'{\i}ctor M. Villalba}
\email{villalba@ivic.ve} \affiliation{Centro de F\'{\i}sica IVIC
Apdo 21827, Caracas 1020A, Venezuela
}%
\author{Clara Rojas}
\email{crojas@ivic.ve} \affiliation{Centro de F\'{\i}sica IVIC Apdo
21827, Caracas 1020A, Venezuela
}%

\begin{abstract}
We solve the Klein-Gordon equation in the presence of a spatially
one-dimensional cusp potential. The scattering solutions are
obtained in terms of Whittaker functions and the condition for the
existence of transmission resonances is derived. We show the
dependence of the zero-reflection condition on the shape of the
potential. In  the low momentum limit, transmission resonances are
associated with half-bound states. We express the condition for
transmission resonances in terms of the phase shifts.
\end{abstract}


\maketitle

The study of low-momentum scattering of nonrelativistic particles by
one-dimensional potentials is a well studied and understood problem
\cite{Newton}. Here we have that, as momentum goes to zero, the
reflection coefficient goes to unity unless the potential $V(x)$
supports a zero energy resonance. In this case the transmission
coefficient goes to unity, becoming a transmission resonance
\cite{Bohm}. Recently, this result has been generalized to the Dirac
equation \cite{Dombey}, showing that transmission resonances at
$k=0$ in the Dirac equation take place for a potential barrier
$V=V(x)$ when the corresponding potential well $V=-V(x)$ supports a
supercritical state. Kennedy \cite{Kennedy} has shown that this
result is also valid for a Woods-Saxon potential. More recently,
transmission resonances and half-bound states have been discussed
for a Dirac particle scattered by a cusp potential
\cite{villalba,Jiang} as well as for a class of short range
potentials \cite{Kennedy2}. The bound states for scalar relativistic
particles satisfying the Klein-Gordon equation  are qualitatively
different from the previous case. Here, for short-range attractive
potentials the Schiff-Snyder effect
\cite{Schiff,Snyder,greiner,Greiner2,Rafelski,Popov,Fulling} takes
place, i.e for a given potential strength two bound states appear,
one with positive norm and another with negative norm. Such states
can be associated with a particle-antiparticle creation process. No
antiresonant states appear \cite{Greiner2,Rafelski}.

The absence of resonant overcritical states for the Klein-Gordon
equation in the presence of short-range potential interactions does
not prevent the existence of transmission resonances for given
values of the potential.

Quantum effects associated with scalar particles in the presence of
external potentials have been extensively discussed in the
literature \cite{greiner,Fulling}. Among quantum effects, we have
that transmission resonance is one of the most interesting
phenomena. For given values of the energy and the proper choice  of
the shape of the effective barrier, the probability of transmission
reaches a maximum such as that obtained in the study of
superradiance \cite{Fulling}, where the amplitude of the scattered
solutions by a rotating Kerr black hole is even larger than the
amplitude of the incident wave. Analogous phenomena can also be
obtained due to the presence of strong electromagnetic potentials
\cite{Manogue}.

Recently, transmission resonances for the Klein-Gordon equation in
the presence of a Woods-Saxon potential barrier  have been computed
\cite{Rojas}. The transmission coefficient as a function of the
energy and  the potential amplitude shows a behavior that resembles
the one obtained for the Dirac equation \cite{Kennedy}. This result
also holds for the square potential \cite{Greiner2}.

In this Letter we discuss the scattering of a Klein-Gordon scalar
particle by the vector cusp potential \cite{villalba}
\begin{equation}
eA^{0}(x)=V(x)=\left\{
\begin{array}{cc}
V_{0}e^{x/a}\quad for\quad x<0, &  \\
V_{0}e^{-x/a}\quad for\quad x>0. &
\end{array}
\right.  \label{V}
\end{equation}
The potential (\ref{V}) vanishes exponentially for large values of
$x$, the parameter $V_{0}$ determines the strength of the barrier
and the constant $a$ defines the width of the potential. The cusp
potential (\ref{V}) reduces
to a repulsive delta interaction of strength $g$ in the limit $%
2aV_{0}\rightarrow g$ as $a\rightarrow 0.$ It is the purpose of the
present Letter to compute the scattering solutions of the
one-dimensional Klein-Gordon equation in the presence of the cusp
vector potential and show that one-dimensional scalar wave solutions
exhibit transmission resonances with a functional dependence on the
shape and strength of the potential similar that  obtained for the
Dirac equation \cite{Kennedy}. The cusp vector potential (\ref{V})
does not possess a square barrier limit and consequently the phase
shift $\delta $, associated with the transmission amplitude, cannot
be directly identified with the positions of the transmission
resonances. \cite{Greiner2}.

The one-dimensional Klein-Gordon equation, minimally coupled to a
vector potential $A^{\mu }$ can be written as
\begin{equation}
\eta ^{\alpha \beta }(\partial _{\alpha }+ieA_{\alpha })(\partial
_{\beta }+ieA_{\beta })\phi +\phi =0,  \label{ws0}
\end{equation}
where the metric $\eta ^{\alpha \beta }=diag(1,-1)$ and here and
thereafter we choose to work in natural units $\hbar =c=m=1$.

Since the potential $V(x)$ in Eq. (\ref{V}) does not depend on time,
we have that $\phi =\phi (x)\exp (-iEt)$, and the problem of solving
the one-dimensional Klein-Gordon equation (\ref{ws0}) reduces to
that of finding solutions to the second-order differential equation
\cite{greiner}
\begin{equation}
\frac{d^{2}\phi (x)}{dx^{2}}+\left[ \left( E-V(x)\right)
^{2}-1\right] \phi (x)=0.  \label{0}
\end{equation}
Let us consider the scattering solutions for $x<0$  with $E^{2}>1$
of the Klein-Gordon equation. We proceed to solve the differential
equation
\begin{equation}
\frac{d^{2}\phi_{L}(x)}{dx^{2}}+\left[ \left( E-V_{0}e^{x/a}\right) ^{2}-1%
\right] \phi_{L}(x)=0.  \label{2}
\end{equation}
On making the variable change $y=2iaV_{0}e^{x/a}$, Eq. (\ref{2})
becomes
\begin{equation}
y\frac{d}{dy}\left( y\frac{d\phi_{L}}{dy}\right) -\left[ \left(
iaE-y/2\right) ^{2}+a^{2}\right] \phi_{L}=0. \label{2a}
\end{equation}

Setting $\phi_{L}=y^{-1/2} f(y)$, Eq. (\ref{2a}) reduces to the
Whittaker equation (\cite{abra}, p. 505)
\begin{equation}
\frac{d^{2}f(y)}{dy^{2}}+\left[ -\frac{1}{4}+\frac{iaE}{y}+\frac{1/4-\mu ^{2}%
}{y^{2}}\right] f(y)=0. \label{a3}
\end{equation}
The general solution of Eq. (\ref{a3}) can be written as
\begin{equation}
\phi _{L}(y)=c_{1}y^{-1/2}M_{\kappa ,\mu }(y)+c_{2}y^{-1/2}M_{\kappa ,-\mu
}(y),
\end{equation}
where $M_{\kappa ,\mu }(y)$ is the Whittaker functions (\cite{abra}
p. 505) and
\begin{equation}
\kappa =iaE,\hspace{1cm}\mu =ia\sqrt{E^{2}-1}.  \label{6}
\end{equation}

Now we consider the solution for $x>0$. In this case, the
differential equation to solve is
\begin{equation}
\frac{d^{2}\phi_{R}(x)}{dx^{2}}+\left[ \left( E-V_{0}e^{-x/a}\right)
^{2}-1\right] \phi_{R}(x)=0.  \label{7}
\end{equation}
On making the variable change $z=2iaV_{0}e^{-x/a}$, Eq. (\ref{7})
can be written as
\begin{equation}
z\frac{d}{dz}\left( z\frac{\phi_{R}}{dz}\right) -\left[ \left(
iaE-z/2\right) ^{2}+a^{2}\right] \phi_{R}=0.
\end{equation}

Putting $\phi_{R}=z^{-1/2} g(z)$ we obtain the Whittaker differential
equation

\begin{equation}
\frac{d^{2}g(z)}{dz^{2}}+\left[ -\frac{1}{4}+\frac{iaE}{z}+\frac{1/4-\mu ^{2}%
}{z^{2}}\right] g(z)=0.
\end{equation}
whose solution is
\begin{equation}
\phi _{R}(z)=c_{3}z^{-1/2}M_{\kappa ,-\mu }(z)+c_{4}z^{-1/2}M_{\kappa ,\mu
}(z),  \label{der}
\end{equation}
Using the asymptotic behavior of the Whittaker function $M_{\kappa
,\mu }(y)$ $\rightarrow e^{-y/2}y^{1/2+\mu }$, as \ $y\rightarrow 0$
(\cite{abra}, p. 504) , we can write the  the incoming wave solution
\ $\phi _{inc}(x)$ in the form

\begin{equation}
\phi _{inc}(x)=c_{1}(2iaV_{0})^{-1/2}e^{-x/{2a}}M_{\kappa ,\mu
}(2iaV_{0}e^{x/a}).
\end{equation}
As $x\rightarrow -\infty $, $\phi _{inc}$ behaves like a plane wave
traveling to the right
\begin{equation}
\phi _{inc}\rightarrow c_{1}(2iaV_{0})^{\mu }e^{i\sqrt{E^{2}-1}x}.
\label{16}
\end{equation}
Analogously, we have that the reflected $\phi _{ref}(x)$ solution can be
written as
\begin{equation}
\phi _{ref}(x)=c_{2}(2iaV_{0})^{-1/2}e^{-x/{2a}}M_{\kappa ,-\mu
}(2iaV_{0}e^{x/a}).
\end{equation}
As $x\rightarrow -\infty ,$ $\phi _{ref}(x)$ \ has the asymptotic behavior
\begin{equation}
\phi _{ref}\rightarrow c_{2}(2iaV_{0})^{-\mu }e^{-i\sqrt{E^{2}-1}x}.
\label{17}
\end{equation}
Finally,  using the right solution $\phi _{R}$ (\ref{der}), we have
that the transmitted solution $\phi _{trans}(x)$ can be expressed as
\begin{equation}
\phi _{trans}(x)=c_{3}(2iaV_{0})^{-1/2}e^{x/{2a}}M_{\kappa ,-\mu
}(2iaV_{0}e^{-x/a}),
\end{equation}
with $c_{4}=0$. As $x\rightarrow \infty $, $\phi _{trans}(x)$ takes the
asymptotic plane wave behavior

\begin{equation}
\phi _{trans}\rightarrow c_{3}(2iaV_{0})^{-\mu }e^{i\sqrt{E^{2}-1}x}.
\label{18}
\end{equation}
The electrical current density for the one-dimensional Klein-Gordon equation
(\ref{ws0}) is given by the expression:
\begin{equation}
j^{\mu }=\frac{i}{2}\left( \phi ^{\ast }\partial ^{\mu }\phi -\phi \partial
^{\mu }\phi ^{\ast }\right)  \label{ws18}
\end{equation}
The current as $x\rightarrow -\infty $ can be decomposed as $%
j_{L}=j_{in}-j_{refl}$ where $j_{in}$ is the incident current and $j_{refl}$
is the reflected one. Analogously we have that, on the right side, as $%
x\rightarrow \infty $ the current is $j_{R}=j_{trans}$, where $j_{trans}$ is
the transmitted current.

Using the reflected $j_{refl}$ and transmitted $j_{trans}$ currents,
we have that the reflection and transmission coefficients $R$ and
$T$ can be expressed as

\begin{equation}
R=\frac{j_{refl}}{j_{inc}},\quad T=\frac{j_{trans}}{j_{inc}}
\end{equation}
The quantities $R$ and $T$ are not independent, they are related via
the unitarity condition $R+T=1$

In order to obtain $R$ and $T$ we proceed to equate at $x=0$ the
right $\phi _{R}$ and left $\phi _{L}$ wave functions and their
first derivatives. From the matching condition we derive the
following  system of equations governing the dependence of
coefficients $c_{1}$ and $c_{2}$ on $c_{3}$. Such set of equations
we solved numerically

\begin{equation}
\label{1a}
 c_{1}M_{\kappa ,\mu}(2iaV_{0})+ c_{2}M_{\kappa
,-\mu}(2iaV_{0})=c_{3}M_{\kappa ,-\mu}(2iaV_{0}),
\end{equation}

\begin{eqnarray}
\label{1b}
c_{1}\left[\left(-\frac{1}{2a}+iV_{0}-\frac{\kappa}{a}\right)M_{\kappa
,\mu}(2iaV_{0})+\frac{1}{a}\left(\frac{1}{2}+\mu+\kappa\right)M_{\kappa+1
,\mu}(2iaV_{0})\right]+\\  \nonumber
+c_{2}\left[\left(-\frac{1}{2a}+iV_{0}-\frac{\kappa}{a}\right)M_{\kappa
,-\mu}(2iaV_{0})+\frac{1}{a}\left(\frac{1}{2}-\mu+\kappa\right)M_{\kappa+1
,-\mu}(2iaV_{0})\right]=\\  \nonumber
=-c_{3}\left[\left(-\frac{1}{2a}+iV_{0}-\frac{\kappa}{a}\right)M_{\kappa
,-\mu}(2iaV_{0})+\frac{1}{a}\left(\frac{1}{2}-\mu+\kappa\right)M_{\kappa+1
,-\mu}(2iaV_{0})\right].
\end{eqnarray}

The reflection
coefficient $R$, and the transmission coefficient $T$, are
calculated by

\begin{equation}
R=\frac{j_{refl}}{j_{inc}}=\frac{\left| c_{2}(2iaV_{0})^{-\mu }\right| ^{2}}{%
\left| c_{1}(2iaV_{0})^{\mu }\right| ^{2}},
\end{equation}
and
\begin{equation}
T=\frac{j_{trans}}{j_{inc}}=\frac{\left| c_{3}(2iaV_{0})^{-\mu }\right| ^{2}%
}{\left| c_{1}(2iaV_{0})^{\mu }\right| ^{2}}.
\end{equation}

From the system of equations (\ref{1a}) and (\ref{1b}) we obtain
that the condition for transmission resonances $T=1$ can be written
as

\begin{eqnarray}
\label{T} \left(\frac{1}{2a}-iV_{0}+\frac{\kappa}{a}\right)
M_{\kappa,\mu}(2iaV_{0})M_{\kappa ,-\mu}(2iaV_{0}) \\ \nonumber
-\frac{1}{2a}\left(\frac{1}{2}+\kappa+\mu\right)M_{\kappa+1,\mu}(2iaV_{0})M_{\kappa
,-\mu}(2iaV_{0}) \\ \nonumber
-\frac{1}{2a}\left(\frac{1}{2}+\kappa-\mu\right)M_{\kappa+1,-\mu}(2iaV_{0})M_{\kappa
,\mu}(2iaV_{0})=0.
\end{eqnarray}

For an attractive cusp potential (\ref{V}), given by $V_{0}<0$, the
bound state solution of the Klein-Gordon equation can be written in
terms of Whittaker functions as \cite{Rojas2}
\begin{equation}
\label{solu}
 \phi_{L}(y)=c_{1}(-y)^{-1/2}M_{-k,\mu}(-y), \hspace{0.5cm}
\phi_{R}(z)=c_{2}(-z)^{-1/2}M_{-k,\mu}(-z),
\end{equation}
where $c_1$ and $c_2$ are normalization constants, and $\phi_{L}$
and $\phi_{R}$ correspond to $x<0$ and $x>0$ respectively. Imposing
the continuity of the solution (\ref{solu}) and of its derivative at
$x=0$, we obtain the bound energy condition
\begin{equation}
[(1+2iaV_{0}-2\kappa )M_{-\kappa \mu }(-2iaV_{0})-(1-2\kappa +2\mu
)M_{{-\kappa +1},\mu }(-2iaV_{0})]M_{-\kappa \mu }(-2iaV_{0})=0.
\label{12}
\end{equation}
In the low momentum limit, the transmission resonance condition
(\ref{T}) reduces to the expression
\begin{equation}
\left[(1-2iaV_{0}+2\kappa )M_{\kappa,0
}(2iaV_{0})-(1+2\kappa)M_{{\kappa +1},0
}(2iaV_{0})\right]M_{\kappa,0 }(2iaV_{0})=0,  \label{Tr}
\end{equation}
It is not difficult to verify that, in the low-momentum limit
$\mu=0$, after the transformation $V_{0}\rightarrow-V_{0}$ and
$E\rightarrow-E$ equation (\ref{Tr}) reduces to the bound energy
condition (\ref{12}) showing that, when the Klein-Gordon equation
exhibits a low-momentum transmission resonance $E=m$ in the presence
of the repulsive cusp potential (\ref{V}), the attractive cusp
potential supports a half-bound state for $E=-m$. This state
corresponds to an antiparticle emerging from the negative-energy
continuum \cite{Greiner2,Rojas2}.

\begin{figure}[th]
\begin{center}
\includegraphics[width=11cm]{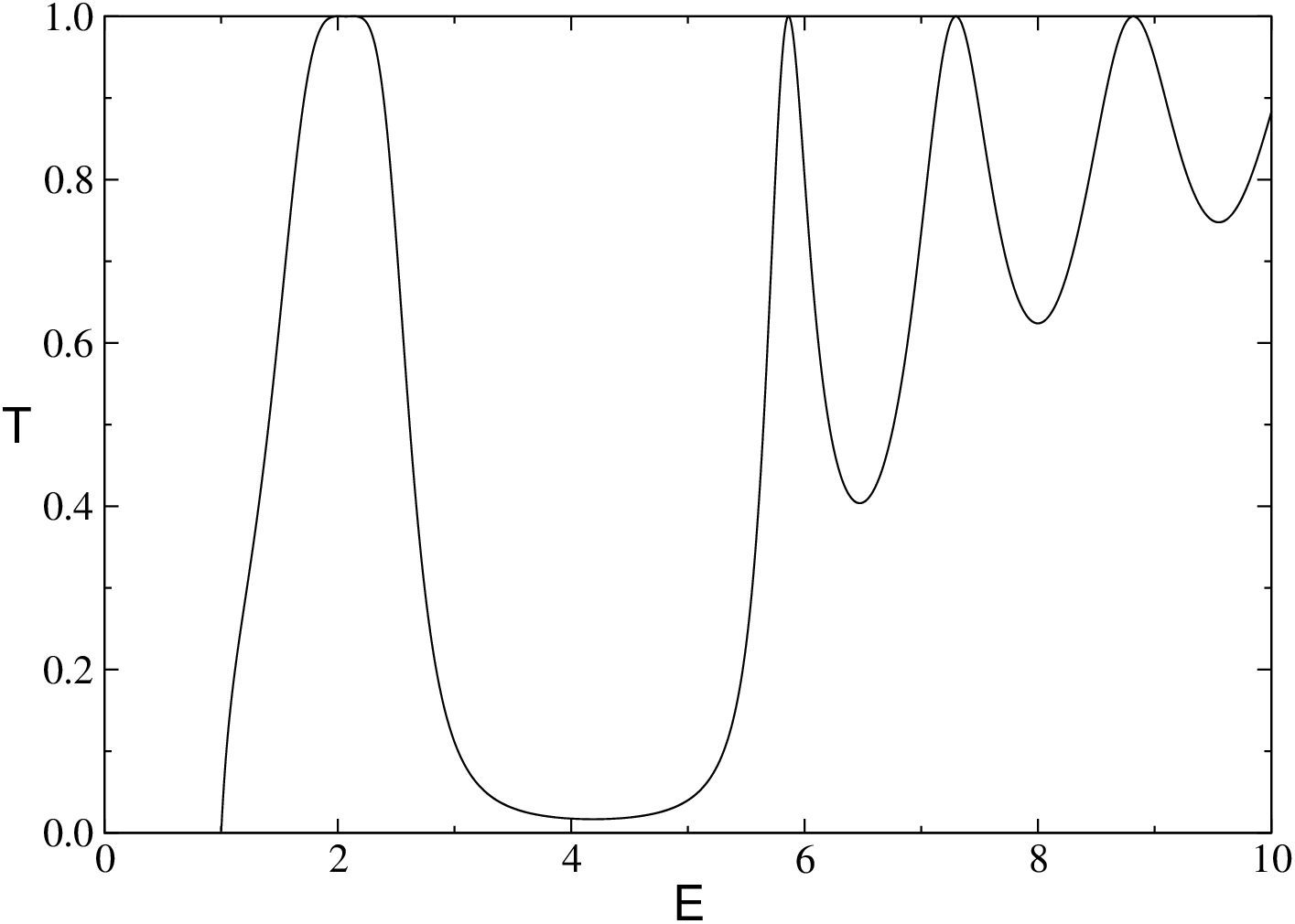}
\caption{Transmission coefficient versus the energy for a square
barrier of height $V_{0}=4$ and width $2L=2$} \label{f11}
\end{center}
\end{figure}

\begin{figure}[th]
\begin{center}
\includegraphics[width=11cm]{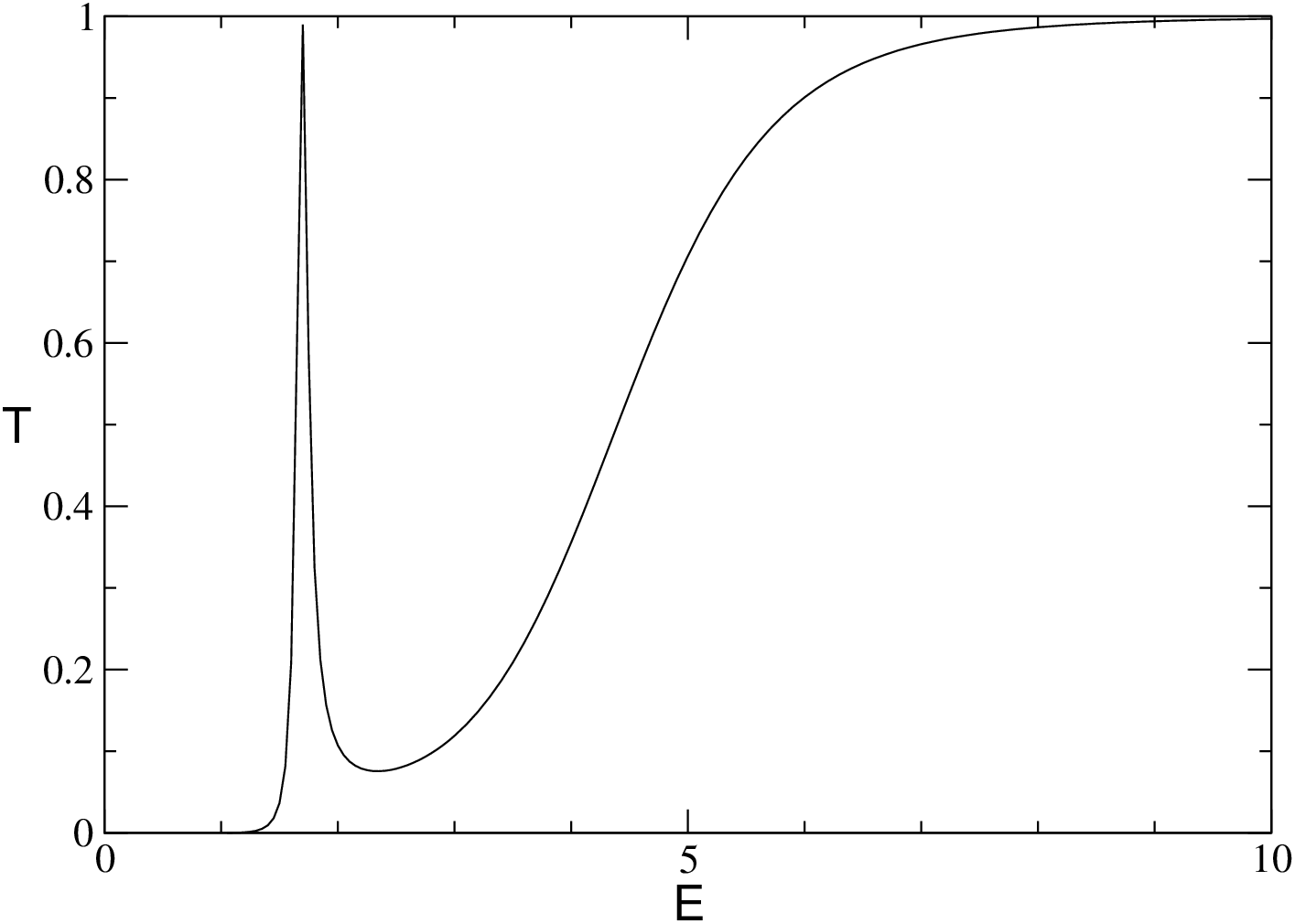}
\caption{The transmission coefficient $T$ versus the energy for
$V_{0}=4$, $a=1$.} \label{f1}
\end{center}
\end{figure}

\begin{figure}[th]
\begin{center}
\includegraphics[width=11cm]{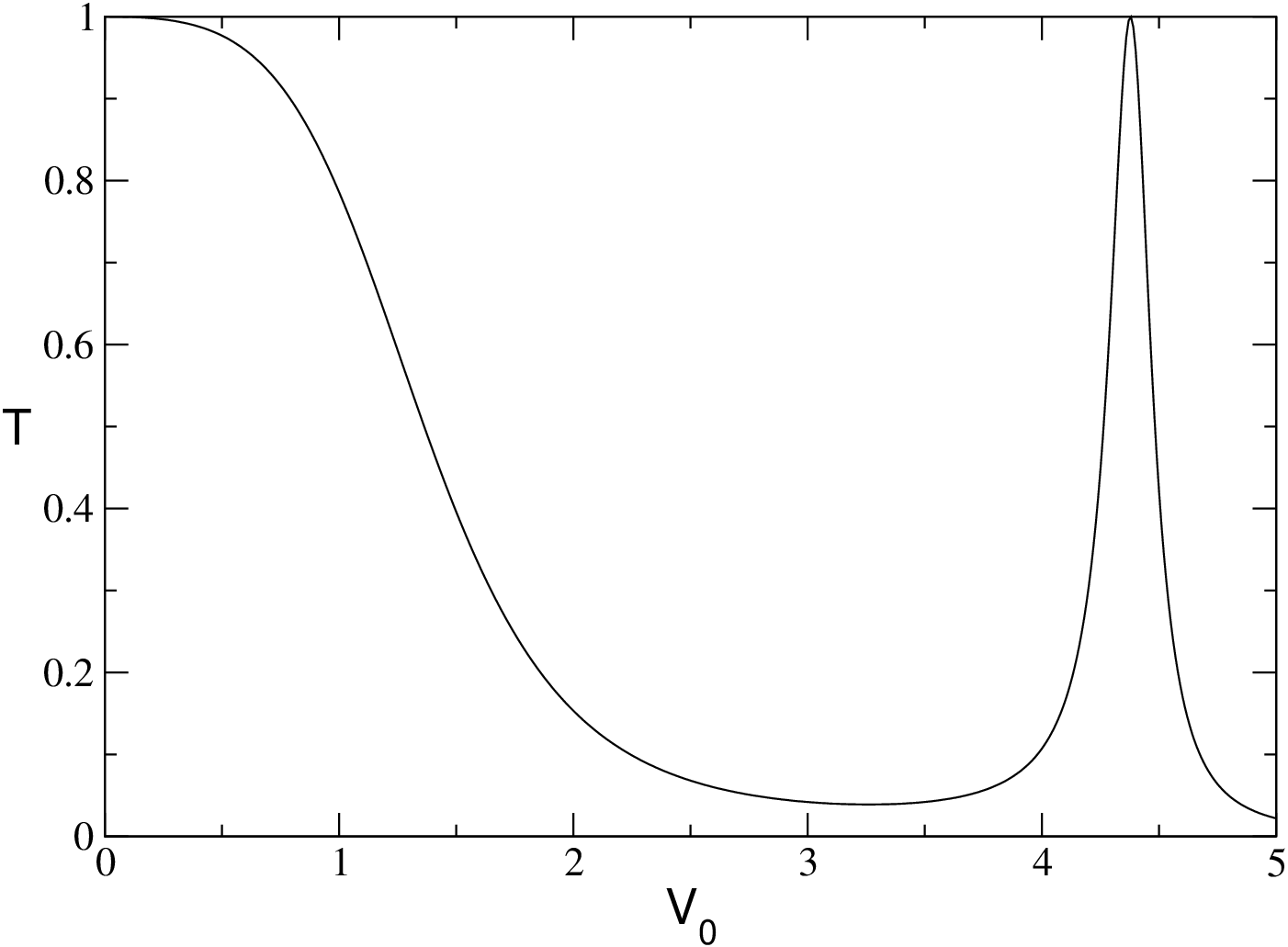}
\caption{The transmission coefficient $T$ versus the potential
strength $V_{0}$ for $E=2$, $a=1$} \label{f2}
\end{center}
\end{figure}
It is instructive to compare the scattering of a scalar Klein-Gordon
particle by a cusp potential with the scattering by a square
barrier. The transmission amplitude for a square barrier of height
$V_{0}$ and width $2L$ is
\begin{equation}
\label{te}
t=\frac{2k_0k_1
e^{-2ik_0L}}{-i(k_0^2+k_1^2)\sin(2k_1L)+2k_0k_1\cos(2k_1L)},
\end{equation}
where
\begin{equation}
\label{k}
 k_0=\sqrt{E^2-1}, \ k_1=\sqrt{(E-V_0)^2-1}.
\end{equation}
Low momentum transmission resonances for a square barrier of height
$V_0$ and width $2L$ can be obtained analytically from equation
(\ref{te}), they satisfy the relation
\begin{equation}
\label{sb}
 \sin(2k_1L)=0, \hspace{0.2cm} \mathrm{with} \hspace{0.2cm} k_0=0.
\end{equation}
From equations (\ref{te})-(\ref{k})  we have that, in the low
momentum limit, a potential well of depth $V_0$ supports half-bound
states at $E=-m$ if $V_0$ satisfies equation (\ref{sb}), as it was
observed for the cusp potential (\ref{Tr}).

Fig. \ref{f11} shows the transmission coefficient as a function of
the energy for a square barrier of height $V_{0}=4$ and width
$2L=2$. It should be noticed that $T(E)$ exhibits a behavior
analogous to the one observed for the Dirac particle by a
Woods-Saxon potential and its square barrier limit \cite{Kennedy}.

From Figures \ref{f1} and \ref{f2}, we can see that, analogous to
the scalar \cite{Rojas} and Dirac \cite{villalba} cases in a
Woods-Saxon potential, the Klein-Gordon equation exhibits
transmission resonances in  the presence of a cusp potential. Figure
\ref{f1} shows that, for $V_{0}=4$ and  $a=1$, a transmission
resonance is reached for $E=1.694$. Figure \ref{f2} shows the
existence of a transmission resonance for $V_{0}=4.378$. It should
be noticed that in both cases the condition $T=1$ is reached for
states with an energy lower the potential strength $V_{0}$. It is
worth mentioning that the cusp potential does not possess a square
well limit. The form of the transmission resonances in Fig. \ref{f2}
shows that, in contrast to the Woods-Saxon potential, there is no
potential strength $V_{0}$ making the cusp barrier completely
impenetrable.

The study of the phase shift $\delta$ of the transmission amplitude
for the cusp potential
\begin{equation}
t=\frac{c_{3}(2iaV_{0})^{-\mu }}{c_{1}(2iaV_{0})^{\mu }}=\sqrt{T}e^{i\delta
}\
\end{equation}
shows that, in contrast to the square barrier potential case for the
Klein-Gordon equation, there is no straightforward relation between
the transmission resonances  and the values for which the phase
shift $\delta $ goes through the value $\pi /2$ or becomes zero. In
fact, using equation (\ref{te}) we have that the phase shift
$\delta$ for the square barrier is
\begin{equation}
\label{delta}
\delta=-2k_0 L
+\arctan\left[\frac{(k_0^2+k_1^2)\tan(2k_1 L)}{2k_0 k_1}\right]
\end{equation}
The condition  for transmission resonances
\begin{equation}
\label{trc}
\sin(2k_{1}L)=0,
\end{equation}
reduces the equation (\ref{delta}) to
\begin{equation}
\delta+2k_0 L=0 \hspace{0.3cm} (\mathrm{mod}\ \pi),
\end{equation}
an analogous result also takes place for  Dirac particle in the
presence of a square well potential (\cite{Greiner2} p. 67).  In
order to get a deeper insight of the nature of the phase shift, and
since cusp potential is even, we proceed to study the scattering
process in terms of solutions of the Klein-Gordon equation (\ref{0})
with a given parity. Using a parity definite basis, we have that the
transmission and reflection amplitudes expressed in terms of the
corresponding phase shifts $\delta _{0}$ and $ \delta _{1}$ are
\cite{Galindo}
\begin{equation}
t=\cos (\delta _{0}-\delta _{1})e^{i(\delta _{0}+\delta _{1})},\
r=\sin (\delta _{0}-\delta _{1})e^{i(\delta _{0}+\delta _{1}+\pi
/2)}. \label{tr}
\end{equation}

From equation (\ref{tr}) one can see that transmission resonances
satisfy the condition

\begin{equation}
\label{do-d1}
 \delta _{0}-\delta _{1}=0 \hspace{0.3cm} (\mathrm{mod}\
\pi)
\end{equation}

For the square barrier we have that $\delta_0-\delta_1$ is
\begin{equation}
\label{dod1}
\delta_0-\delta_1=\arctan\left[2\frac{\left(k_1^2-k_0^2\right)}{k_0k_1}%
\sin\left(2k_1L\right)\right],
\end{equation}
the transmission resonance condition for the square barrier
(\ref{trc}) reduces the equation (\ref{dod1}) to the condition
(\ref{do-d1}).

\begin{figure}[th]
\begin{center}
\includegraphics[width=11cm]{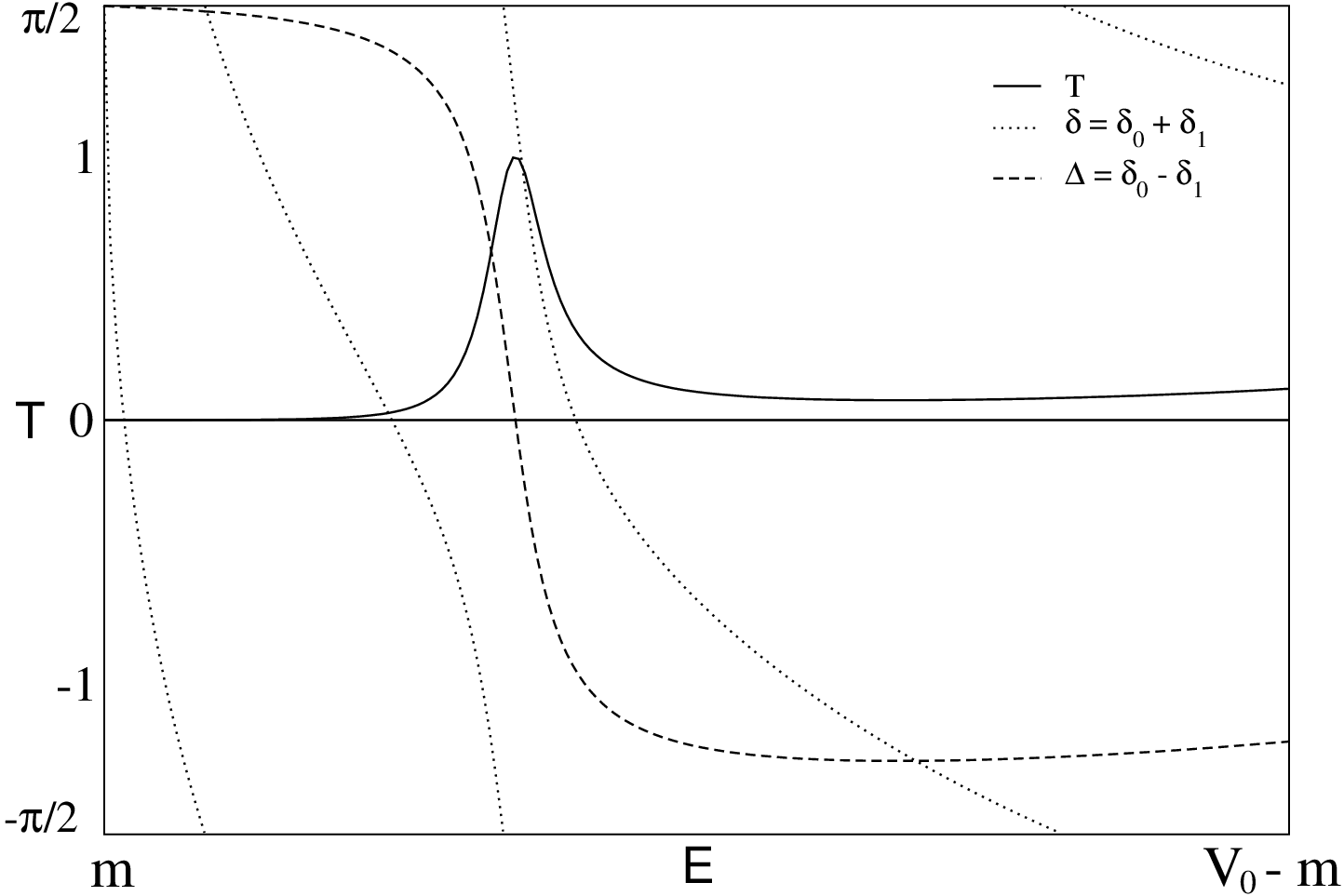}
\caption{Dependence of the phase shifts
$\delta=\delta_{0}+\delta_{1}$ and $\Delta=\delta_{0}-\delta_{1}$ on
the energy for $V_{0}=4$ and $a=1$. The condition $T=1$ is reached
at $\Delta=0$ $(\mathrm{mod} \ \pi)$  \label{f6}}
\end{center}
\end{figure}

The analytic computation of the phase shifts $\delta_0$ and
$\delta_1$ for the cusp potential (\ref{V}) involves very cumbersome
expressions, nevertheless, looking at the expressions in Eq.
(\ref{tr}), it becomes clear that for arbitrary values of $V_{0}$
and $a$, the transmission resonances in the cusp potential (\ref{V})
cannot be directly associated with the phase shift $\delta=\delta
_{0}+\delta _{1}$. Since the phase of the reflected wave is
$\delta+\pi/2$, the knowledge of the phase shifts does not suffice
for computing the transmission resonances of the Klein-Gordon
equation in the presence of the cusp potential. Fig \ref{f6} shows
the dependence of the transmission resonances on the phase shifts
$\delta_{0}$ and $\delta_{1}$.

\begin{figure}[th]
\begin{center}
\includegraphics[width=11cm]{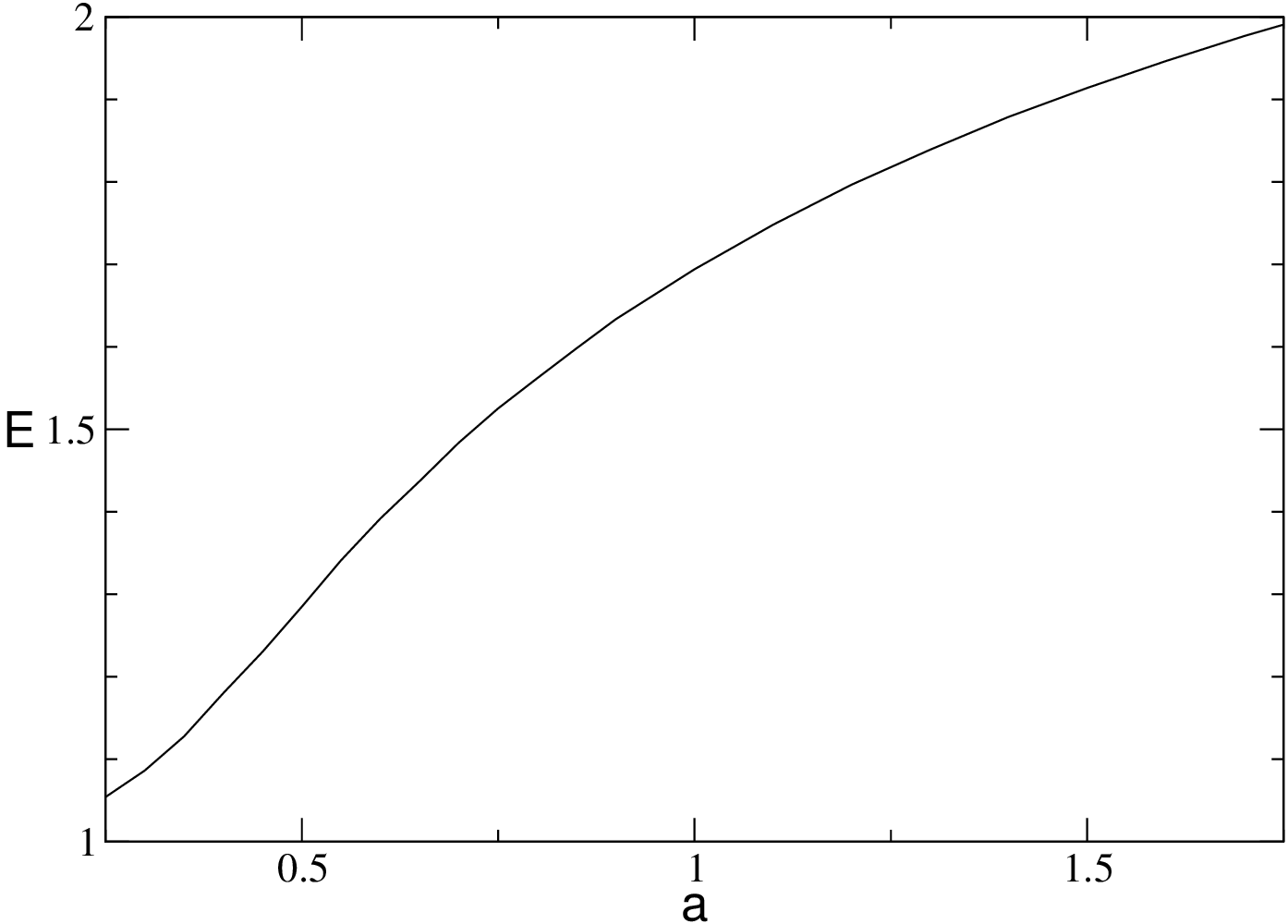}
\caption{Dependence of the transmission resonance energy on the
potential shape $a$ for $V_{0}=4$. Notice that the energy for which
$T=1$ decreases as the potential cusp becomes sharper.} \label{f4}
\end{center}
\end{figure}

\begin{figure}[th]
\begin{center}
\includegraphics[width=11cm]{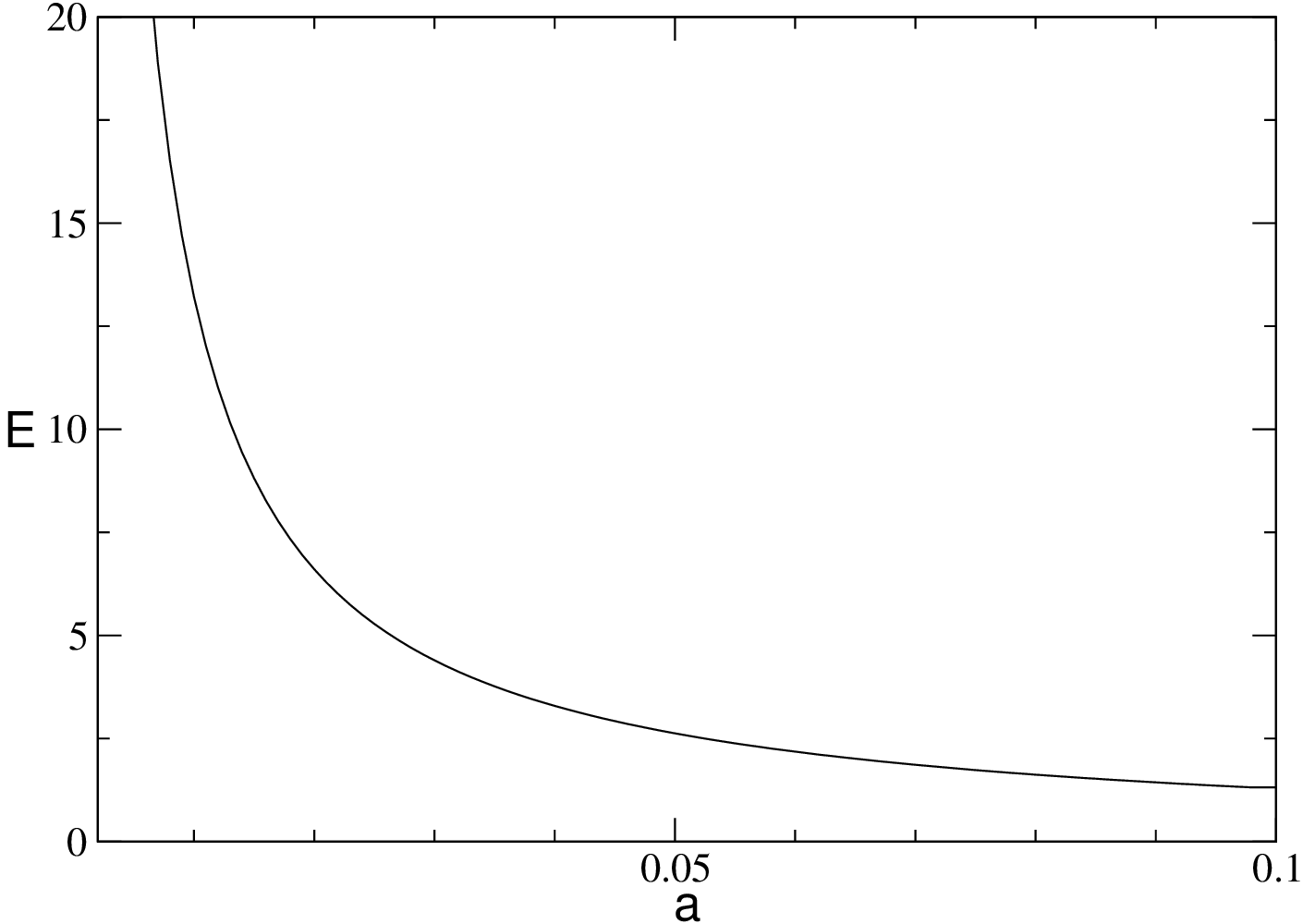}
\caption{Dependence of the transmission resonance energy on the
shape parameter $a$ for a cusp potential with $2aV_{0}=1$, notice
that the energy increases as the potential becomes sharper}
\label{f5}
\end{center}
\end{figure}

Fig. \ref{f4} shows that, for a given value $V_{0}$ of the potential
strength, the transmission resonance energy decreases as the cusp
becomes sharper.  The point interaction delta limit is illustrated
in Fig. \ref{f5}. In this case we observe that, as the potential
cusp approaches to the Dirac delta, the energy necessary for a
transmission resonance increases, diverging in the limit
$a\rightarrow 0$.

From equation (\ref{delta}) we have that,  in the low momentum
limit, the phase shift $\delta$ for the square barrier becomes zero
if the potential strength satisfies the transmission resonance
condition (\ref{trc}),  otherwise the phase shift takes the value
$\delta=\pi/2$. Looking at the system of equations (\ref{1a}) and
(\ref{1b}), we obtain that this statement also holds for the cusp
potential (\ref{V}). Fig. \ref{f4} shows that, for a given value
$V_{0}$, not satisfying the condition (\ref{Tr}), the transmission
coefficient vanishes in the low momentum limit and the phase
$\delta$ reaches the value $\delta= \pi/2$.

We have obtained that, in the low-momentum limit, the transmission
coefficient of Klein-Gordon particles incident on  the cusp
potential (\ref{V}) is $T=1$ provided that the potential supports a
half-bound state at $E=-m$. We have also shown that, analogous to
the Dirac particle scattered  by a square barrier \cite{Dombey},
low-momentum transmission resonances of scalar particles are
associated with supercritical states. Nevertheless, for the square
well and the attractive cusp potential, half-bound states correspond
to antiparticle states that join the bound particle state and they
both disappear to create a neutral condensate \cite{Greiner2}.

In conclusion, in this Letter we have shown that the Klein-Gordon
equation exhibits transmission resonances for potentials that do not
possess a square-barrier limit. We have seen that, despite the fact
that the supercritical behavior of the bound states in the
one-dimensional cusp potential \cite{Popov,Rojas2} is qualitatively
different from the one observed for Dirac particles \cite{villalba},
transmission resonances possess the same structure observed for the
Dirac equation.

\acknowledgments{We thank Dr. Ernesto Medina for reading and
improving the manuscript. This work was supported by FONACIT under
project G-2001000712.}

\end{document}